# Synthesis, Crystal Structure and Physical Properties of $Sr_2FeOsO_6$

Avijit Kumar Paul,[a] Martin Jansen,[a]*

Binghai Yan,[a] Claudia Felser,[a] Manfred Reehuis,[b] and Paula M. Abdala[c]

[a] Max Planck Institute for Chemical Physics of Solids, Nöthnitzer Straße 40, Dresden 01187, Germany

[b] Helmholtz-Zentrum Berlin für Materialien und Energie, D-14109 Berlin, Germany

[c] SNBL at ESRF, BP 220, F-38042 Grenoble Cedex 9, France

*Corresponding author: Max Planck Institute for Chemical Physics of Solids, Nöthnitzer Straße 40, Dresden 01187, Germany

E-mail address: Martin.Jansen@cpfs.mpg.de

**ABSTRACT**

In the exploration of new osmium based double perovskites, $Sr_2FeOsO_6$ is a new insertion in the existing family. The polycrystalline compound has been prepared by solid state synthesis from the respective binary oxides. PXRD analysis shows the structure is pseudo-cubic at room temperature, whereas low-temperature synchrotron data refinements reveal the structure to be tetragonal, space group $I4/m$. Heat capacity and magnetic measurements of $Sr_2FeOsO_6$ indicated the presence of two magnetic phase transitions at $T_1 = 140$ K and $T_2 = 67$ K. Band structure calculations showed the compound as a narrow energy gap semiconductor, which supports the experimental results obtained from the resistivity measurements. The present study documents significant structural and electronic effects of substituting $Fe^{3+}$ for $Cr^{3+}$ ion in $Sr_2CrOsO_6$.

## 1. INTRODUCTION

The discovery of high tunneling magneto resistance (TMR) at room temperature in $Sr_2FeMoO_6$[1] has drawn great attention to the field of double perovskites of general formula $A_2BB'O_6$ with $A$ an alkaline earth such as Ca, Sr or Ba and $B$, $B'$ two different transition metals. Indeed, exploring the double perovskite systems more extensively has revealed a wide spectrum of interesting physics. The most important properties encountered are colossal magnetoresistance, half metallic ferrimagnetism, high $T_C$ ferrimagnetism and multiferroicity.[1-7] This impressive spectrum of exciting phenomena, as well as predictions of exotic electronic and magnetic structures of various double perovskites[8,9] have motivated us to further explore materials in this potentially quite extensive family.

Strong spin-orbit coupling in 4$d$- and 5$d$-transition metals introduces anomalies in magnetic and electronic properties in double perovskites hosting such elements. Recently, 5$d$-based double perovskites have been brought to the focus of attention due to their high magnetic ordering temperatures ($T_C$) and complex exchange mechanisms originating from spin-orbit coupling.[2] Among the double perovskites with the highest $T_C$ is $Sr_2CrReO_6$ with $T_C$ = 635 K.[10] Our recent investigations on $Sr_2CrOsO_6$ have revealed it to be an unprecedent high $T_C$ (725 K) ferrimagnetic insulator.[11] This conspicuous effect has let it appear attractive to optimize it by varying the 3$d$ element. Although several representatives of 3$d$ and 4$d$ (5$d$) double perovskites like $Sr_2FeMO_6$ ($M$ = W, Mo, Ru, Re) were reported earlier,[12] respective Os and Ir containing species have not been studied in detail. We realized that $Sr_2FeOsO_6$ might be a promising candidate in the context of solids displaying competing magnetic interactions.

Recent theoretical studies on $Sr_2FeOsO_6$ have suggested that the compound would show ferrimagnetic ordering at 65 K,[8] and the energetically most stable structure would be the tetragonal variant in space group $I4/m$.[13] An early experimental study by Sleight et al. reported $Sr_2FeOsO_6$ to be cubic.[14] However, no further investigations on structure and properties have become available for this material till date. Here we report on experimental data featuring strong magnetic exchange couplings, and on comparisons of earlier predicted properties. In detail, we have synthesized phase pure polycrystalline $Sr_2FeOsO_6$, refined its crystal structure from synchrotron powder data at different temperatures and have performed basic physical characterizations such as heat capacity, transport and magnetic properties measurements.

## 2. EXPERIMENTAL SECTION

**2.1. Synthesis of $Sr_2FeOsO_6$.** The title compound was synthesized as polycrystalline powder from stoichiometric amounts of binary oxides at 1273 K in a sealed quartz tube under argon atmosphere. The reagents, $SrO_2$ (Aldrich, 99%) and $OsO_2$ (Sigma Aldrich, 83% Os) were used as received and without any further purifications. $Fe_2O_3$ was freshly prepared by thermal decomposition of iron acetate (Sigma Aldrich, 95%) in a flow of oxygen at 400°C. A typical batch consisted of a stoichiometric mixture of $SrO_2$ (0.239 g), $Fe_2O_3$ (0.079 g) and $OsO_2$ (0.222 g), which was ground thoroughly inside a glovebox. The starting mixture was pressed into pellets that were placed in corundum containers and finally sealed in quartz tubes under argon atmosphere. Pure single phase polycrystalline $Sr_2FeOsO_6$ was obtained after 50 hours of heating at 1273 K. The heating and cooling rates were kept at 50 K/hour, throughout. In order to assure full oxygen occupancies, the product was reheated at 773 K for two days in a gold crucible under 120 MPa oxygen pressures, in a steel autoclave.

**2.2. X-ray powder diffractometry.** Laboratory X-ray powder diffraction (XRPD) studies at room temperature were performed with a D8-Advance diffractometer (Bruker AXS, Karlsruhe, Germany) Mo-Kα radiation (λ = 0.71073 Å), coverning a 2θ range of 5 - 45 Rietveld refinements were carried out with the program TOPAS.[15]

Synchrotron powder patterns were recorded at 15 and 78 K with the high-resolution powder diffractometer on the BM1B (Swiss Norvegian) beam line at the ESRF in Grenoble. The 2θ- range covered was 2 to 50 ° using synchrotron radiation with the wavelength λ = 0.50357(2) Å. For this experiment, the sample was filled and sealed in a 0.5 mm diameter glass capillary. Rietveld refinements of the powder diffraction data were carried out with the program *FullProf* using atomic scattering factors provided by the *FullProf* program.[16]

**2.3. Physical Measurements.** The magnetization was measured in the temperature range from 2 to 350 K using a Quantum design MPMS-XL7 SQUID magnetometer. The electric resistance was determined on sintered polycrystalline pellets (diameter 5 mm and thickness 1 mm) by the ordinary four-probe method in the temperature range from 25 to 300 K. Temperature dependence of the specific heat ($C_p$) of polycrystalline $Sr_2FeOsO_6$ was measured between 2 and 300 K with a commercial PPMS (Physical Property Measurement System, Quantum Design, 6325 Lusk Boulevard, San Diego, CA.). To thermally fix the sample tablet (30 mg) to the sapphire sample platform, a minute amount of Apiezon-N vacuum grease was used. The heat capacity of the sample holder platform and grease was individually determined in a separate run and subtracted from the total measured heat capacities.

**2.4. Theoretical calculations**.

Density-functional theory calculations were performed within the generalized gradient approximation (GGA)[17] and GGA+U levels,[18] which are implemented into the VASP package[19]. The projector-augmented-wave potential[20] was employed to represent the core electrons. An 8×8×8 k-point mesh was adopted for the Brillouin zone sampling in the total energy integral. All the atomic parameters used for the calculations were the experimental ones.

**3. RESULTS AND DISCUSSION**

**3.1. Crystal structure of $Sr_2FeOsO_6$.** Figure 1 shows the laboratory X-ray powder diffraction pattern taken at room temperature. Our first trials on laboratory data showed that the crystal structure could be successfully refined in the cubic space group $Fm\bar{3}m$, resulting in the lattice parameter $a$ = 7.8591(1) Å. A better fit was achieved, $R_{wp}$= 0.067 dropping to $R_{wp}$= 0.061, by assuming minor disorder of Fe and Os cations over the $B(B')$ sublattice positions. The partial occupancy of Os at the Fe sites was found to be 6.0(1) %. From our synchrotron data recorded at lower temperature we obtained a partial occupation of 4.3(2) %. The cubic crystal structure is shown in the inset of Figure 1 and the refined structural parameters are summarized in Table 1. The observed Fe-O, Os-O and Sr-O bond lengths and bond angles are comparable to other similar double perovskites.[21-23]

In order to check for a potential change from the cubic to a lower symmetric crystal structure we collected data sets at the lower temperature. Two synchrotron powder patterns were collected at 78 K and 15 K (Figure S1 and Figure 2). In the powder pattern collected at 15 K (Figure 2), a pronounced splitting was observed for the cubic reflection (400) at a diffraction angle of about 14.7°. Here the ratio of the peak intensities is close to 1: 2, suggesting the presence of a tetragonal splitting. Further it has to be mentioned that no additional peaks were observed in the low-temperature pattern, confirming the *I*-centering of the crystal structure. For ordered double perovskites $A_2BB'X_6$ group theoretical methods have been used to deduce possible lower-symmetric structures reflecting octahedral tiltings.[24,25] We describe the crystal structure in the tetragonal space group *I*4/*m* (No. 87) with the cell dimensions ½($a + b$) × ½($a - b$) × $c$ in reference to the cubic cell.[25] The space group *I*4/*m* (No. 87), admits tilting around the *c*-axis (tilt system $a^0a^0c^-$).[26] In this space group the atoms are located at the following positions: Sr in 4$d$(0,½,¼), Fe in 2$a$(0,0,0), Os in 2$b$(0,0,½), O1 in 4$e$(0,0,$z$), and O2 in 8$h$($x,y$,0); $x$(O2) and $y$(O2) can thus be refined independently. The refinement of the crystal structure in *I*4/*m* resulted in a satisfactory residual $R_F = 0.025$ (defined as $R_F = \sum ||F_{obs}|-|F_{calc}||/\sum |F_{obs}|$), but a weighted $\chi^2$ value of 2.41. The resulting bond lengths are: $d$(Fe-O1) = 2.000(7) Å and $d$(Fe-O2) = 1.971(7) Å; $d$(Os-O1) = 1.969(7) Å and $d$(Os-O2) = 1.962(7) Å. It is interesting to compare the Fe-O bond distances with those values found for the low-temperature monoclinic ferrite CaFeO$_3$, where one has observed a charge-disproportionated phase containing $Fe^{3+}$ and $Fe^{5+}$ ions[27] and cooperative Jahn-Teller elongation was found to be absent, as one would expect for ions in 3$d^5$ and 3$d^3$ configurations. From our data it can be concluded that the metal ions in Sr$_2$FeOsO$_6$ seem to be close to the nominal $Fe^{3+}$ and $Os^{5+}$ valence states.

After analysing the low-temperature data, re-inspection of room temperature XRPD data revealed a significant facet of the present structure. A careful analysis of the XRPD shows the diffraction peaks at 20.7° and 33.1° were significantly broadened, reflecting presence of a tetragonal unit cell. Re-refinement of the PXRD data with *I*4/*m* symmetry resulted in better figures of merit ($R_{wp}$= 0.049, $R_p$= 0.033). All the respective refined parameters and bond lengths are summarized in Table 1, which are in well accordance with the low temperature data. So, the room temperature structure is found as pseudo-cubic, and should be prone to transform into a cubic structure at elevated temperature.

**3.2. Structural Comparison**

In the crystal structures of $Sr_2MOsO_6$ (*M* = Cr and Fe), $MO_6$ and $OsO_6$ corner sharing octahedra are alternatingly arranged, while Sr ions are coordinated by 12 oxide ions. Both, the Fe and Cr representatives, are forming closely related double perovskite structures. In the present study, we have observed a tetragonal distortion for $Sr_2FeOsO_6$ at low temperature (Figure 3). In contrast, $Sr_2CrOsO_6$ transforms from cubic to rhombohedral. Such structural transitions can be classified employing Glazer's notation, addressing symmetry reduction.[25] Glazer proposed a simple method for assigning space groups to perovskites, considering all possible patterns of octahedral tiltings, starting from the cubic aristotype. P. Woodward has extended this approach to the double perovskites $A_2BB'O_6$.[28] According to Glazer's notation, double perovskite structures with the space groups $Fm\bar{3}m$, *I*4/*m* and $R\bar{3}$ are described as $a^0a^0a^0$, $a^0a^0c^-$ and $a^-a^-a^-$, respectively. In the present system, two Fe-O1-Os angles are 180° along the *c* axis and four Fe-O2-Os angles are 165.3° in the *ab* plane (at 15K). In $Sr_2CrOsO_6$, all six Cr-O-Os connections enclose an angle of 172.4° (at 2 K). The O-Cr-O and O-Os-O angles deviate from the values of perfect octahedral (90° and 180°) angles. From a structural point of view, two kinds of deviations from an ideal double perovskite structure characterize the rhombohedral structure, whereas the tetragonal structure type shows one type of distortion along the Fe-O2-Os bonds (adjacent angle of two octahedra in *ab* plane).

The stability of perovskites $ABO_3$ depends on the ionic radii of the constituting metal ions, which is generally assured by the tolerance factor *t* (equ. 1).

$$t = \frac{r_A + r_O}{\sqrt{2}(r_B + r_O)} \quad \cdots\cdots (1)$$

$r_A$, $r_B$, and $r_O$ are the ionic radii of the respective ions. In double perovskites $A_2BB'O_6$, $r_B$ is the average of $r_B$ and $r_B$. For an ideal cubic structure, the value of t is equal to or near unity. The tolerance factors for Sr$_2$MOsO$_6$ with $M$ = Cr and Fe are obtained to be 1.007 and 0.999, respectively. It is known for rhombohedral symmetry that $t > 1.00$, whereas $1 > t > 0.97$ directs to tetragonal symmetry. Although the deviations discussed are minute, they correctly reflect the trend observed experimentally, see Figure 3. The slight deviations of tolerance factors from $t = 1$ also correspond to the observed tendency to adopt the cubic structure at RT, or slightly above.

### 3.3. Magnetic properties

In Sr$_2$FeOsO$_6$, we consider Fe and Os ions to be in a trivalent and pentavalent state corresponding to electron configurations of 3d$^5$ for Fe$^{3+}$ and 5d$^3$ for Os$^{5+}$, respectively. Performing magnetization measurements as a function of temperature and magnetic field strength we have probed the local and collective magnetic responses. The temperature dependence of the magnetic susceptibility for Sr$_2$FeOsO$_6$ in applied fields of 0.1, 1 and 10 kG is shown in Figure 4a. Zero-field cooled (ZFC) and field cooled (FC) data were collected over the temperature range 2─350 K. Magnetic susceptibility data at 0.1 kG field shows the presence of two maxima at 140 and 67 K, indicating antiferromagnetic ordering with two Néel temperatures. Fitting the high-temperature susceptibility ($250 < T < 350$) to the Curie-Weiss law, given by

$$\chi_{CW} = C/(T - \theta)$$

where C is the Curie constant, and $\theta$ is the Curie-Weiss temperature, results in values of $\mu_{eff}$ = 4.32 $\mu_B$, C = 2.33 emu/K mol$^{-1}$ and $\theta$ = +80 K. The observed average magnetic moment is much lower than expected for a spin-only contribution (Fe$^{3+}$: $d^5$, S = 5/2; Os$^{5+}$: $d^3$, S = 3/2) which we attribute to spin-orbit coupling, as it was found in other osmium double perovskite as well.[21-23] Positive value of $\theta$ implies dominant ferromagnetic interactions in the paramagnetic region. With lowering the temperature below 250 K, the paramagnetic response is substantially enhanced until at $T_{N1}$ = 140 K the plot shows downturn behavior. This first ordering at 140 K ($T_{N1}$) appears to be suppressed with increasing the applied magnetic field, while antiferromagnetic ordering at 67 K remains unchanged with varying the field strength. The divergences between ZFC-FC curves extend over a wide temperature range (Figure 4a). Significant deviations between ZFC and FC is still present below 67 K ($T_{N2}$), which would be in conflict with an ideal ordered antiferromagnetic structure.

The field dependencies of the magnetization are shown in Figure 4b. Well above the transition temperature ($T_{N2}$) the plot is virtually linear (at 77K), indicating the ideal antiferromagnetic behavior. However, the field dependencies are no longer linear and show an unexpected hysteresis at 5 K. The saturation magnetization is observed to amount only to about 0.34 $\mu_B$, which is very low in comparison to expectations for a ferromagnetic ground state. The magnetic properties as reflected by the $\chi(T)$ data are thus of intriguing complexity. With decreasing temperature, the plot exhibits an antiferromagnetic downturn of the susceptibility at 140 and 67 K, despite the fact that the positive Weiss constant obtained for the temperature range 250 – 350 K suggests ferromagnetic interactions. Below $T_{N2}$, the title compound displays canted antiferromagnetism, and ZFC-FC splitting indicates presence of magnetic frustration. Recently, zur Loye et al. reported on $Sr_2NiOsO_6$ and $Sr_2CuOsO_6$, which are antiferromagnetic with Weiss constants of +27 and −40 K for Ni and Cu, respectively.[21,22] However, in these compounds, the Os ions are in a hexavalent state whereas Os is pentavalent in $Sr_2FeOsO_6$.

In comparison to $Sr_2CrOsO_6$, the present compound shows a quite peculiar magnetic behavior and a surprisingly low ordering temperature. According to the Goodenough-Kanamori rules, both the $d^5$-O-$d^3$ combinations Fe-O-Cr and Fe-O-Os are supposed to exhibit FM coupling. In the title compound, Fe-O-Os is found to be AFM in the *ab* plane while it is FM along the *c* axis. We argue that the local structure of Fe-O-Os bonds can be crucial for the super-exchange interaction. Along the *c* axis the Fe-O-Os angle is 180º, which favors the FM coupling. However, the Fe-O-Os angle varies to 165.3º in the *ab* plane, which may explain the emerging AFM coupling. In addition, Os-d states are at higher energy, show a larger $e_g$-$t_{2g}$ gap and stronger spin-orbit coupling than those of Cr-d states, which will also affect the super-exchange coupling. Since both the FM and AFM couplings are effective simultaneously, the competition between these two different mechanisms can reduce the stability of the current magnetic configuration resulting in much lower magnetic transition temperature.

### 3.4. Transport properties

The resistivity was measured from 300 to 25 K. The temperature dependence of the electric resistivity of the slowly cooled sample is shown in Figure 5. The compound shows an increase of the resistivity with decreasing temperature, and thus is semiconducting with the temperature variation of the resistivity being approximately Arrhenius-like. Since the investigated samples are polycrystalline, grain boundary effects cannot be ruled out. In spite of this limitation, we trust that we have obtained evidence for the correct qualitative conducting property of the oxide under investigation. The activation energy of the conduction was revealed from the Arrhenius plot (inset Figure 5.) to 125 meV, which is comparable with other semiconducting double perovskites.[4]

### 3.5. Heat capacity measurements

The specific heat for Sr2FeOsO6 was recorded in the temperature range of 2 to 300 K (Figure 6). In the low temperature region one can see a λ-type anomaly at 67 and a broad hump at 140 K in the $C_P(T)$ curve, which we assign to the magnetic transition temperatures discussed above. To probe the nature of the specific heat anomaly at $T_{N's}$ in more detail, we plotted $C_P/T$ vs. $T$ as shown in Figure 6, allowing to determine the Néel temperatures more precisely. While the first transition temperature (140 K) gives a broad hump, the second transition shows a sharp, λ-type peak at 67 K, which suggests on- set of long range magnetic ordering.

### 3.6. Electronic Structure

For the first-principle calculation, we adopted the atomic structures and tetragonal lattice parameters measured at low temperature (15K), in which a = 5.5174 Å and c = 7.9389 Å. The primitive unit cell includes two $Sr_2FeOsO_6$ formula units, as shown in Figure 7. We checked different magnetic configurations of Fe and Os sites in our total energy calculations and found that the antiferromagnetic (AFM) structure shown in Fig. 6 is the most stable one. As shown in the density of states (DOS) in Figure 8, GGA calculation exhibits a small energy gap of 0.06 eV. The employment of Hubbard U can effectively increase the energy gap. For example, U = 4 eV for Fe-3$d$ states and U = 2 eV for Os-5$d$ states increase the gap to 0.73 eV. Employing U can also push Fe-3$d$ states into deep bands. Since the calculated gap by GGA+U is much larger than the experimental value of 0.125 eV, we expect that the effective U of Fe-3$d$ states should be smaller than 4 eV. Here, we present results with $U_{Fe-3d}$ = 0 and 4 eV as two limits of the band structure. In addition, spin-orbit-coupling is found to reduce the energy gap slightly by 0.1 eV through broadening Os-5$d$ states.

Concerning the AFM magnetic structures, our analyses have revealed neighboring Fe1 and Os1 (see Figure 7) to couple within the $ab$ plane of the tetragonal lattice in an AFM way, with Fe1 having a larger magnetic moment than Os1. As an example, Fe1 (Os1) presents magnetic moments of 3.7 (-1.3) $\mu_B$ in GGA calculations. This value is found to increase to 4.2 (-2.1) $\mu_B$ applying GGA+U. On the other hand, neighboring Fe and Os sites couple with each other in a ferromagnetic (FM) way along the $c$ axis. Equivalently, Fe (Os) sites from neighboring planes exhibit opposite spin polarization, resulting in a net zero magnetic moment. From the projected DOS, we obtain the $d$-orbital occupation, as illustrated in Fig. 7. The Fe ion is in the $Fe^{3+}$ ($3d^5$) state, while Os is in the $Os^{5+}$ ($5d^3$) state. Five spin-up electrons occupy the $e_g$-$t_{2g}$ levels of Fe1, and three spin-down electrons occupy the $t_{2g}$ levels of Os1, while its $e_g$ level is empty. Consequently, the lowest conduction bands are constituted by $t_{2g}$ states of Fe and $e_g$ states of Os, which are found to hybridize with each other in DOS (see Figure 8). These lowest conduction bands get narrower when decreasing $c/a$ (tetragonal elongation), since both Fe-$t_{2g}$ and Os-$e_g$ states split less in this case.

## 4. CONCLUSIONS

In the course of our systematic investigations aiming at the discovery of structurally and physically interesting materials, we have presented our findings on another osmium based double perovskite, $Sr_2FeOsO_6$. Samples with full occupancies on the oxygen sites, warranted by post-annealing under elevated oxygen pressure, exist as low temperature tetragonal and room temperature pseudo-cubic polymorphs. This constitutes a contrast to the chromium analogue which exists at low temperature in a rhombohedrally distorted double perovskite structure. Specific-heat and SQUID measurements showed effects at 140 and 67 K, indicating onsets of long-range magnetic ordering. Although the crystal structures of the Fe and Cr representatives differ but slightly, interestingly the transition temperatures of the iron compound were found to be much lower than that of $Sr_2CrOsO_6$. Investigations of electronic conductivity characterize the title compound as a semiconductor. The electronic structure calculations have confirmed that the compound is a narrow energy gap semiconductor. Energetically most stable at low temperature is a tetragonal structure in *I*4/*m* with antiferromagnetic magnetic ordering. Our study provides scope and optimism to synthesize new double perovskites with employing different combinations of 3*d*-5*d* transition metals, which are prone to exhibit special chemical and physical properties. The present investigations along with our earlier studies[11] clearly suggest that the chemistry of the Os-based double perovskite is rich and provide avenues for further research. A detailed discussion of the magnetic structure requires more information about the individual contributions from spin and orbital magnetic moments of Fe and Os as may be obtained from neutron diffraction experiments.

**ACKNOWLEDGEMENTS**

We thank G. Siegle for assisting the conductivity and specific heat measurements. We also thank R. K. Kremer and E. Brucher for supporting the magnetic measurements. A. K. Paul wants to acknowledge specially O. V. Magdysyuk for fruitful discussion of TOPAS.

**Supporting Information**: Supporting materials available include Synchrotron powder diffraction collected at 78 K (S1); separated portions of all powder diffractions (S2-S4); XRPD pattern at RT using tetragonal unit cell (S5-S6).

**REFERENCES**

(1) Kobayasi, K.-I.; Kimura, T.; Sawada, H.; Terakura, K.; Tokura, Y. *Nature (London)*, **1998**, *395*, 677.

(2) Serrate, D.; Teresa, J. M. Da.; Ibarra, M. R. *J. Phys.: Condens. Matter* **2007**, *19*, 023201.

(3) Dass, R. I.; Goodenough, J. B. *Phys. Rev. B* **2003**, *67*, 014401.

(4) (a) Mandal, T. K.; Gopalakrishnan, J. *Chem. Mater*. **2005**, *17*, 2310. (b) Nag, A.; Manjanna, J.; Tiwari, R. M.; Gopalakrishnan, J. *Chem. Mater*. **2008**, *20*, 4420.

(5) Dass, R. I.; Yan, J.-Q.; Goodenough, J. B. *Phys. Rev. B* **2004**, *69*, 094416.

(6) Gao, H.; Liobet, A.; Barth, J.; Winterlik, J.; Felser, C.; Panthofer, M.; Tremel, W. *Phys. Rev. B* **2011**, *83*, 134406.

(7) Felser, C.; Fecher, G. H.; Balke, B. *Angew. Chem. Int. Ed*. **2007**, *46*, 668.

(8) Meetei, O. N.; Erten, O.; Randeria, M.; Trivedi, N.; Woodward, P. *Phys. Rev. Let*., **2013**, *110*, 087203.

(9) Mandal, T. K.; Felser, C.; Greenblatt, M.; Kubler, J. *Phys. Rev. B* **2008**, *78*, 134431.

(10) Kato, H.; Okuda, T.; Okimato, Y.; Tomioka, Y.; Takenoya, Y.; Ohkubo, A.; Kawasaki, M.; Tokura, Y. *Appl. Phys. Lett.* **2002**, *81*, 328.

(11) (a) Krockenberger, Y.; Mogare, K.; Reehuis, M.; Tovar, M.; Jansen, M.; Vaitheeswaran, G.; Kanchana, V.; Bultmark, F.; Delin, A.; Wilhelm, F.; Rogalev, A.; Winkler, A.; Alff, L. *Phys. Rev. B* **2007**, *75*, 20404. (b) Krockenberger, Y.;Reehuis, M.; Tovar, M.; Mogare, K.; Jansen, M.; Alff, L. *J. Magn. Magn. Mat*. **2007**, *310*, 1854.

(12) Retuerto, M.; Jimenez-Villacorta, F.; Martinez-Lope, M. J.; Huttel, Y.; Roman, E.; Fernandez-Diaz, M. T.; Alonso, J. A. *Phys. Chem. Chem. Phys.,* **2010**, *12*, 13616.

(13) Wang, J.; Meng, J.; Wu, Z. *Chem. Phys. Lett.* **2011**, *501*, 324.

(14) Sleight, A.W.; Longo, J.; Ward, R. *Inorg. Chem*., **1962**, *1*, 245.

(15) *TOPAS 3.0* by Bruker AXS GmbH Karlshruhe, Germany.

(16) Rodríguez-Carvajal, J. *FullProf*: a Program for Rietveld Refinement and Pattern Matching Analysis, Abstract of the Satellite Meeting on Powder Diffraction of the XV Congress of the IUCr, p. 127, Toulouse, **1990**.

(17) Perdew, J. P.; Burke, K.; Ernzerhof, M. *Phys. Rev. Lett*., **1996**, *77*, 3865.

(18) Liechtenstein, A. I.; Anisimov, V. I.; Zaanen, J. *Phys. Rev. B,* **1995**, *52*, R5467.

(19) (a) Kresse, G.; Hafner, J. *Phys. Rev. B* **1993**, *47*, 558. (b) Kresse, G; Furthmüller, J. *Phys. Rev. B* **1996**, *54*, 11169.

(20) Kresse, G.; Joubert, D. Phys. Rev. B 1999, 59, 1758.

(21) Tian, C.; Wibowo, A. C.; zur Loye, H.-C.; Whangbo, M. H. *Inorg. Chem*., **2011**, *50*, 4142.

(22) Macquart, R.; Kim, S. J.; Gemmill, W. R.; Stalick, J. K.; Lee, Y.; Vogt, T.; zur Loye, H.-C. *Inorg. Chem*., **2005**, *44*, 9676.

(23) Gemmill, W. R.; Smith, M. D.; Prozorov, R.; zur Loye, H.-C. *Inorg. Chem*., **2005**, *44*, 2639.

(24) Howard, C. H.; Kennedy, B. J.; Woodward, P. M. *Acta. Cryst*. **2003**, *B59*, 463.

(25) Bock, O.; Müller, U. *Acta. Cryst.* **2002**, *B58*, 594.

(26) Glazer, A. M. *Acta. Cryst.* **1972**, *B28*, 3384.

(27) Woodward, P. M.; Cox, D. E.; Moshopoulou, E.; Sleight, A. W.; Morimoto, S. *Phys. Rev. B* **2000**, *62*, 844.

(28) Woodward, P. M. *Acta. Cryst.* **1997**, *B53*, 32.

**Table 1:** Results of the crystal structure refinements of $Sr_2FeOsO_6$ as obtained from laboratory *x*-ray (at RT) and synchrotron data (at 15 and 78 K). In the cubic structure with $Fm\bar{3}m$ the atoms are in the following positions: Sr in $8c$(¼,¼,¼), Fe in $4a$(0,0,0), Os in $4b$(½,½,½), and O in $24e$(¼,0,0); In the tetragonal one with $I4/m$: Sr in $4d$(0,½,¼), Fe in $2a$(0,0,0), Os in $2b$(0,0,½), O1 in $4e$(0,0,$z$), and O2 in $8h$($x,y$,0), respectively.

| | **PXRD at RT** | **PXRD at RT** | **Synchrotron at 78 K** | **Synchrotron at 15 K** |
|---|---|---|---|---|
| Crystal system | Cubic | Tetragonal | Tetragonal | Tetragonal |
| Space group | $Fm\bar{3}m$ (No. 225) | $I4/m$ (No. 87) | $I4/m$ (No. 87) | $I4/m$ (No. 87) |
| $a$ (Å) | 7.8591(1) | 5.5485(1) | 5.5227(1) | 5.5174(1) |
| $c$ (Å) | 7.8591 | 7.8867(1) | 7.9291(1) | 7.9389(1) |
| $V$ (Å$^3$) | 485.433(1) | 242.806(1) | 241.838(2) | 241.673(2) |
| Distortion ($c/\sqrt{2}a$) | — | 1.005 | 1.015 | 1.017 |
| $R_{wp}$ (%) | 6.09 | 4.94 | 10.3 | 12.4 |
| $R_{exp}$ (%) | 2.31 | 2.31 | 7.76 | 9.73 |
| $R_p/R_F$ (%) | 3.89 | 3.28 | 2.92 | 2.51 |
| $R_{bragg}$ (%) | 3.24 | 3.85 | 4.16 | 4.13 |
| $x$(O) | 0.25 | $x$(O1) = 0.00<br>$x$(O2) = 0.2495(1) | $x$(O1) = 0.00<br>$x$(O2) = 0.2815(11) | $x$(O1) = 0.00<br>$x$(O2) = 0.2827(11) |
| $y$(O) | 0.00 | $y$(O1) = 0.00<br>$y$(O2) = 0.2491(1) | $y$(O1) = 0.00<br>$y$(O2) = 0.2205(10) | $y$(O1) = 0.00<br>$y$(O2) = 0.2184(10) |
| $z$(O) | 0.00 | $z$(O1) = 0.2455(1)<br>$z$(O2) = 0.00 | $z$(O1) = 0.2487(8)<br>$z$(O2) = 0.00 | $z$(O1) = 0.2519(8)<br>$z$(O2) = 0.00 |
| $B_{eq}$(O) | 0.855(1) | 0.132(1) | 0.686(81) | 0.761(86) |
| $d_{(Fe-O)}$ (Å) | 1.964(1) | $d_{(Fe-O1)}$ = 1.936(1)<br>$d_{(Fe-O2)}$ = 1.939(1) | $d_{(Fe-O1)}$ = 1.972(8)<br>$d_{(Fe-O2)}$ = 1.975(8) | $d_{(Fe-O1)}$ = 2.000(7)<br>$d_{(Fe-O2)}$ = 1.971(7) |
| $d_{(Os-O)}$ (Å) | 1.964(1) | $d_{(Os-O1)}$ = 2.007(1)<br>$d_{(Os-O2)}$ = 1.985(1) | $d_{(Os-O1)}$ = 1.993(8)<br>$d_{(Os-O2)}$ = 1.959(8) | $d_{(Os-O1)}$ = 1.969(7)<br>$d_{(Os-O2)}$ = 1.962(7) |

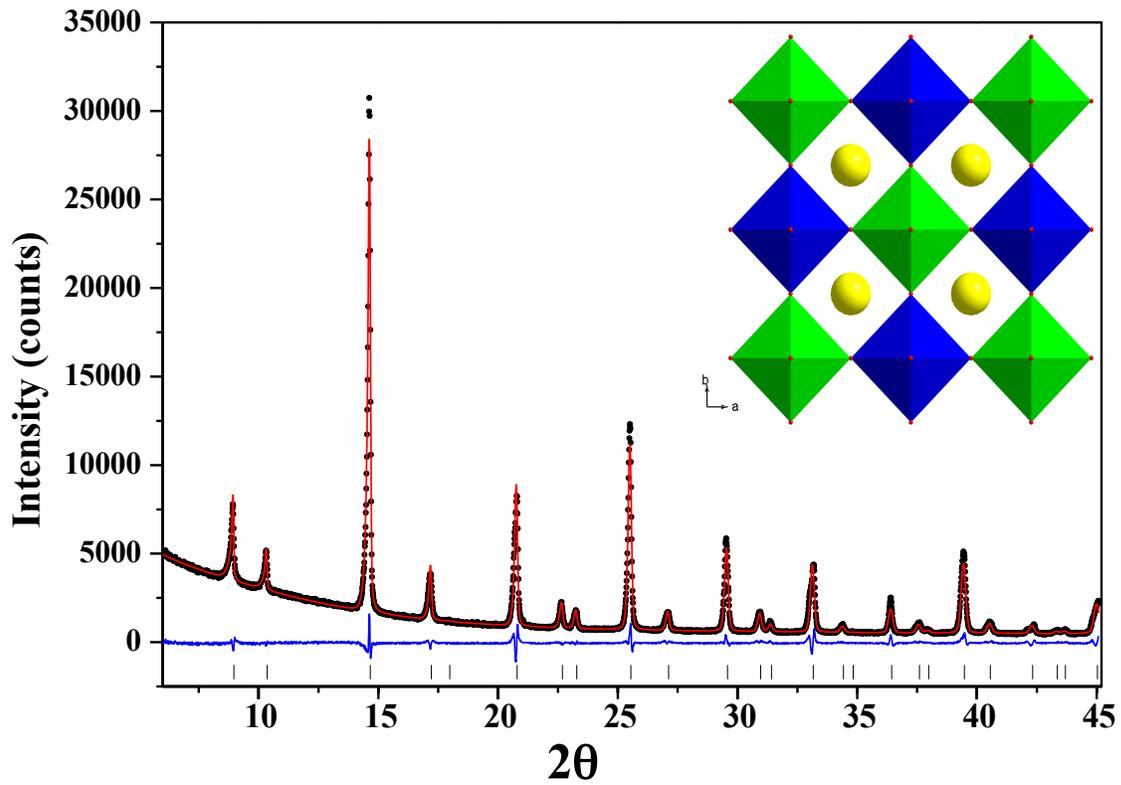

**Figure 1**. Room temperature X-ray powder diffraction pattern of $Sr_2FeOsO_6$ (black spheres, observed; red line, fit from Reitveld refinement in $Fm\bar{3}m$ ; blue line, difference curve; lower black bars, Bragg peaks;). Inset shows the crystal structure of the compound (yellow spheres, Sr; red spheres, O; green octahedral, Fe; blue octahedral, Os).

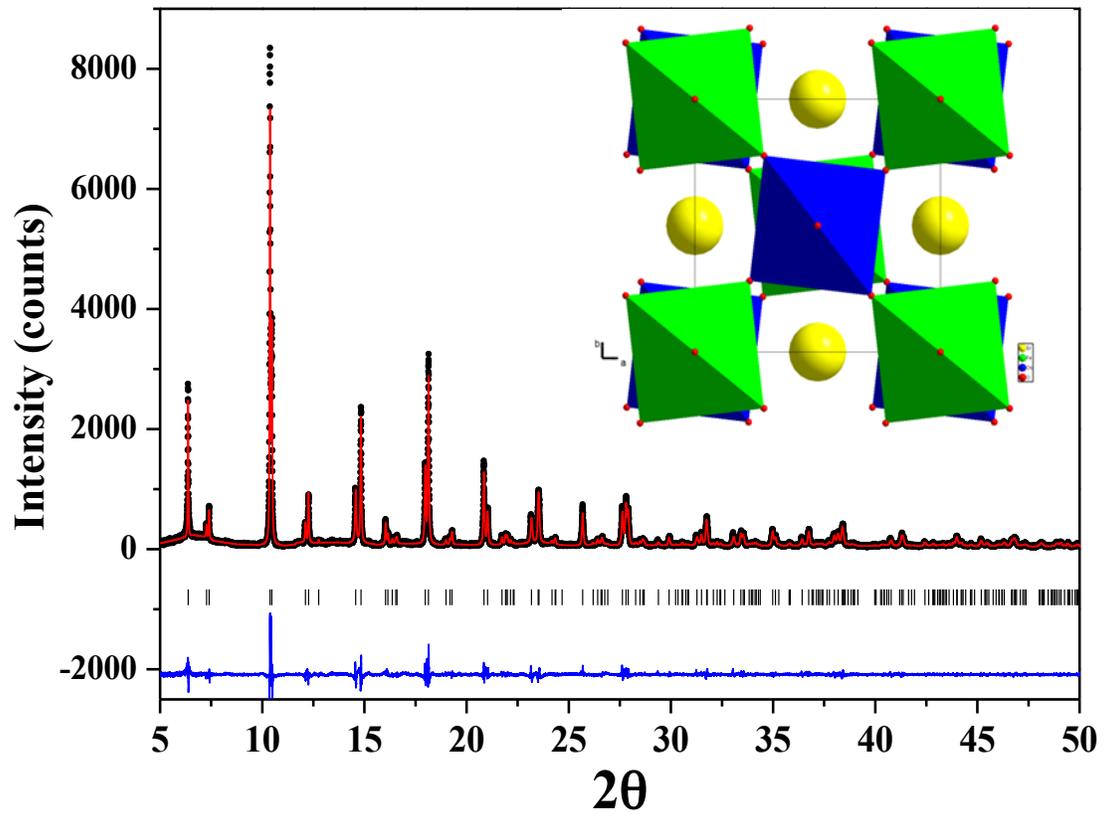

**Figure 2**. Synchrotron diffraction pattern of $Sr_2FeOsO_6$ at 15 K (black spheres, observed; red line, fit from Reitveld refinement; black bars, Bragg peaks; lower blue line, difference curve;). Inset shows the crystal structure of the compound (yellow spheres, Sr; red spheres, O; green octahedral, Fe; blue octahedral, Os).

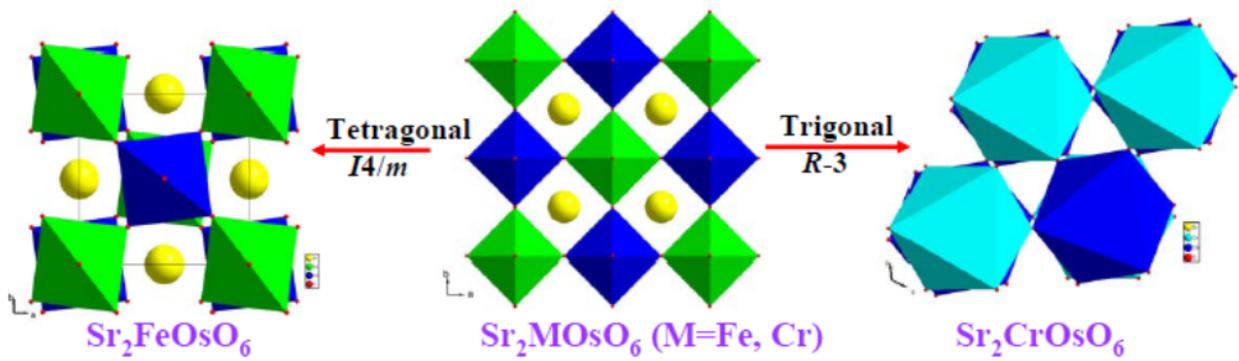

**Figure 3**. Structural diagram for phase transition in Fe and Cr based double perovskites. $Sr_2FeOsO_6$ shows the tetragonal distortion and $Sr_2CrOsO_6$ shows the trigonal distortion at lower temperature.

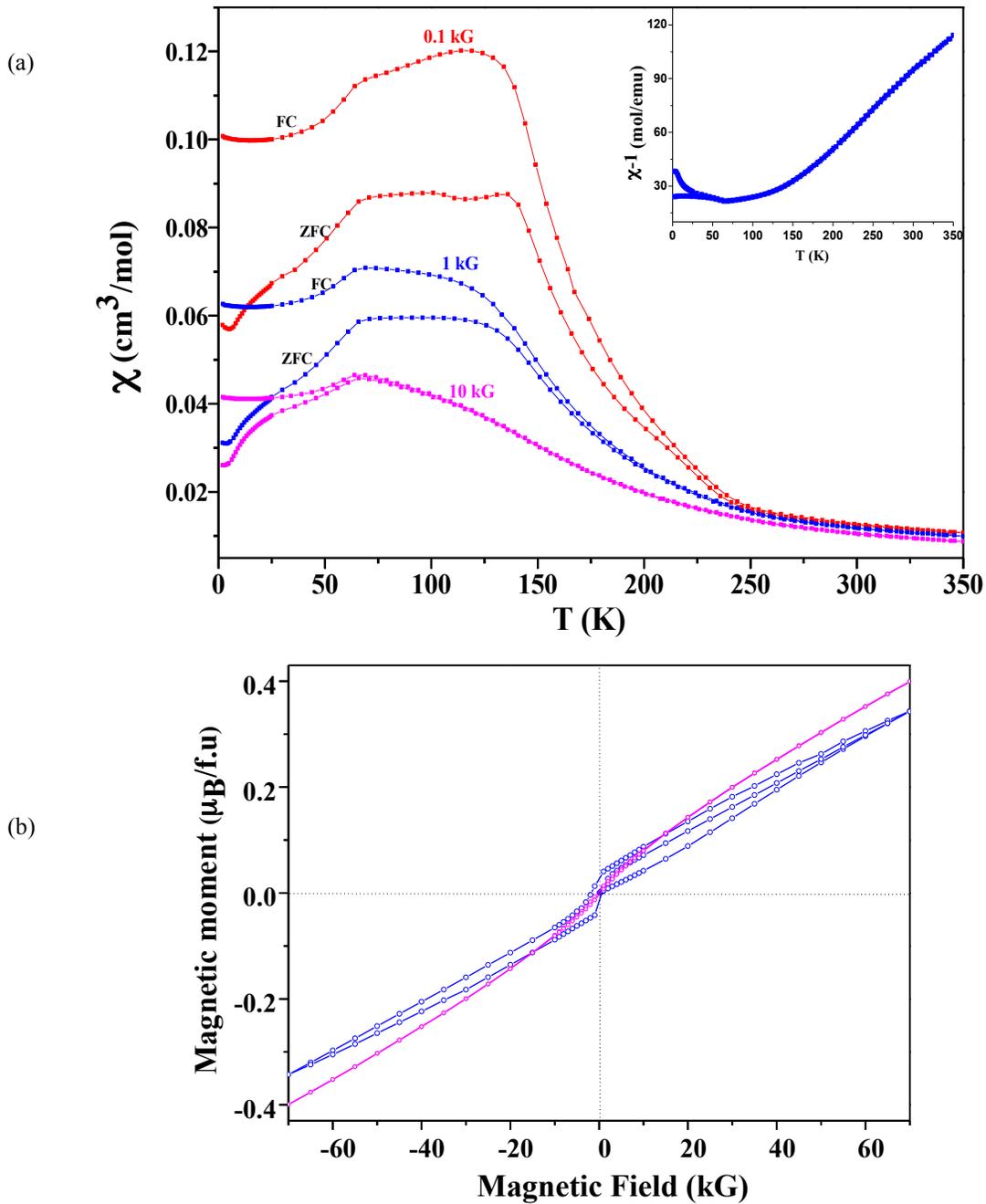

**Figure 4**. (a) Temperature dependence of the ZFC-FC susceptibilty of $Sr_2FeOsO_6$ in different applied field. Red points correspond to 0.1 kG field; Blue points correspond to 1 kG field; Pink points correspond to 10 kG field. Inset shows the inverse susceptibility plot with the variation of temperature at 10 kG field. (b) Field dependence of the magnetization measured at different temperature. Blue line corresponds to 5 K and pink line corresponds to 77 K.

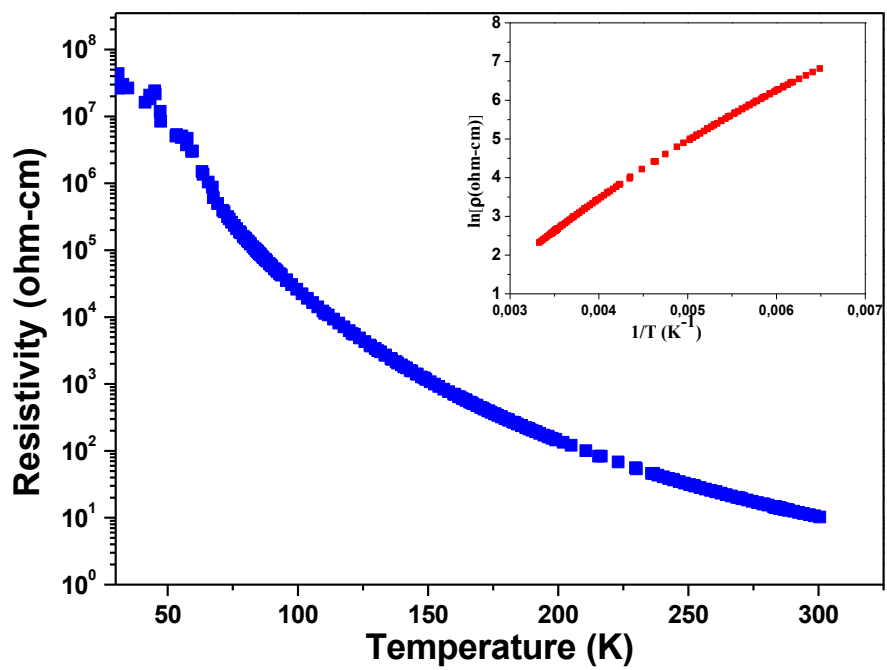

**Figure 5**. Temperature dependence of the resistivity of $Sr_2FeOsO_6$. Inset shows the Arrhenius fit.

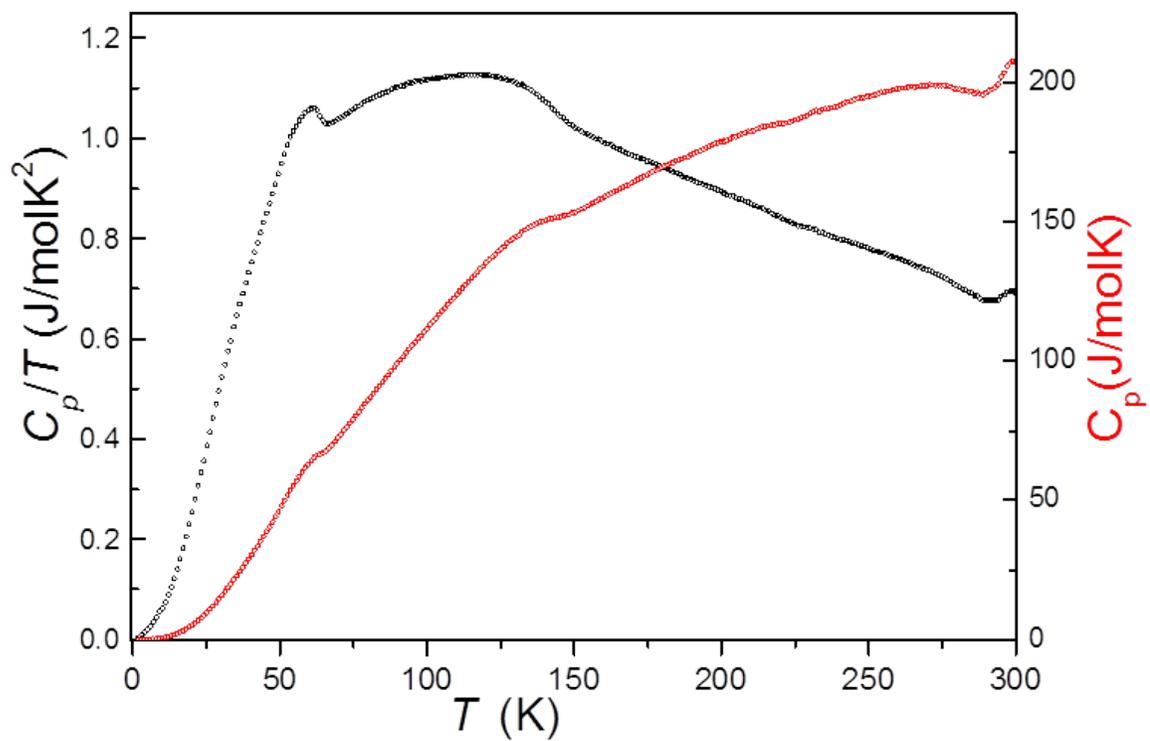

**Figure 6**. Heat capacity measured as a function of temperature.

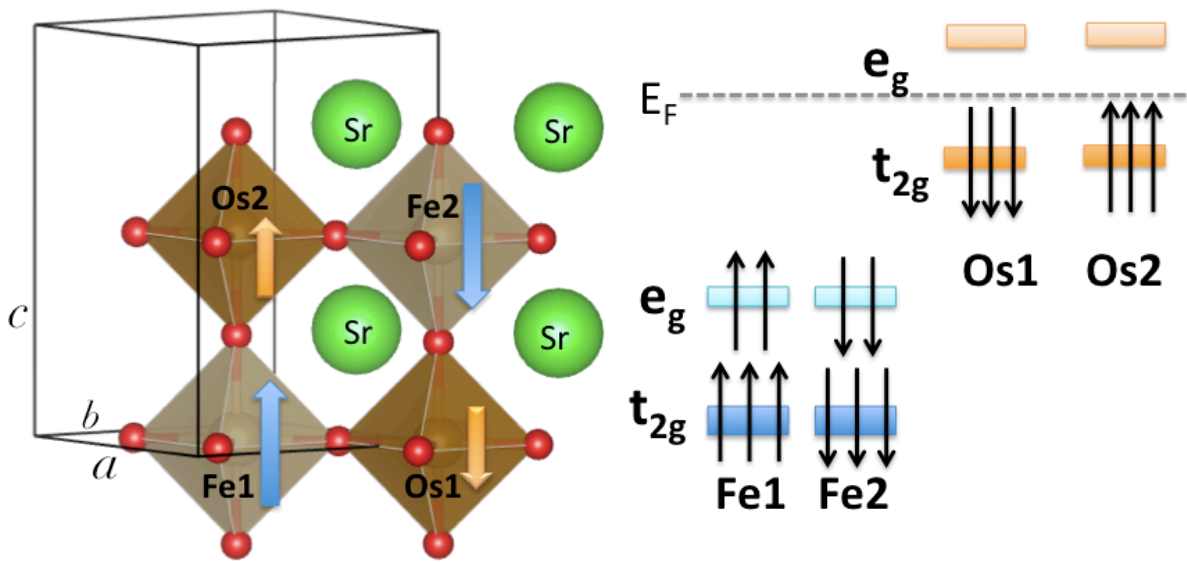

**Figure 7**. Tetragonal lattice structure of $Sr_2FeOsO_6$. Large arrows indicate the magnetic moments of Fe1, Fe2, Os1 and Os2. The red balls represent O atoms. The *a,b* and *c* axes are labeled for the tetragonal lattice, where the tetragonal box indicates the primitive unit cell. The *d*-orbital occupations of Fe ($Fe^{3+}$, $3d^5$) and Os ($Os^{5+}$, $5d^3$) sites are illustrated on the right.

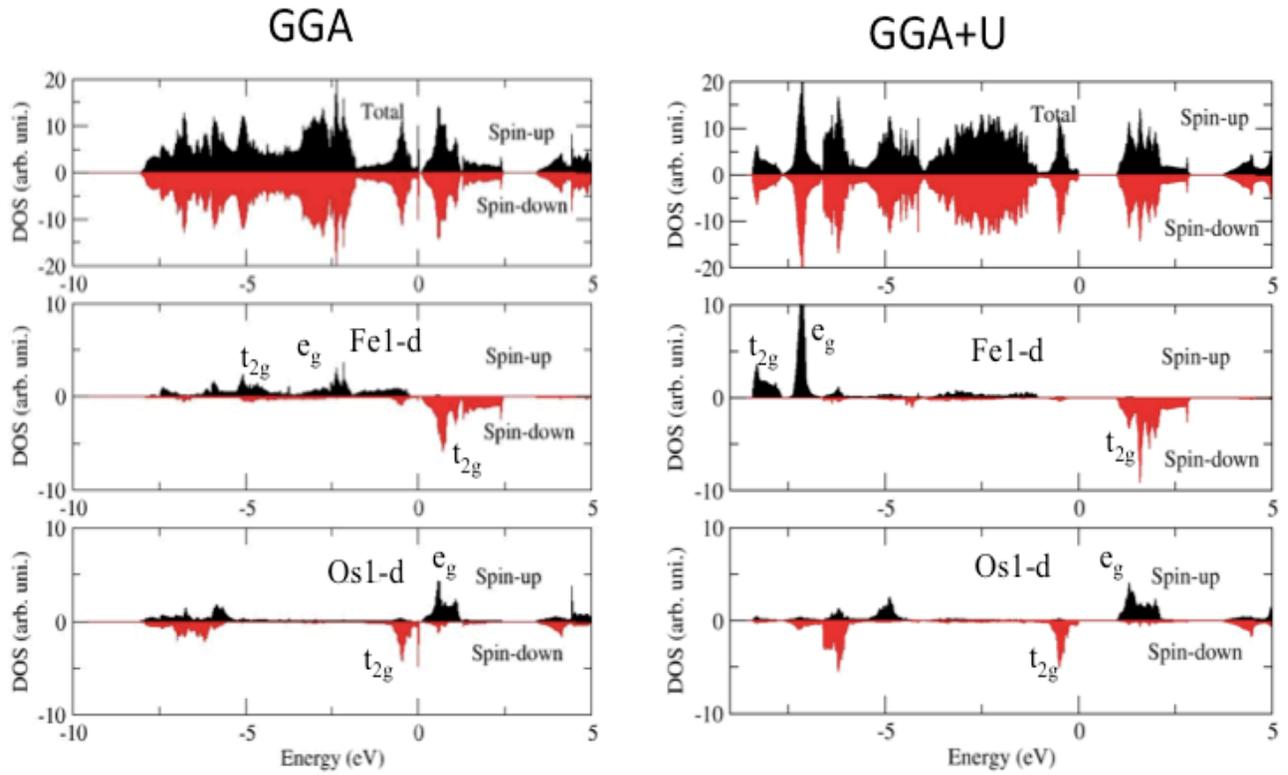

**Figure 8**. DOS of $Sr_2FeOsO_6$ from GGA and GGA+U calculations. The Fermi energy is shifted to zero. Spin-up and down states are indicated by positive (black) and negative (red) values. Since Fe2 (Os2) is known to present an opposite spin-polarization to Fe1 (Os1), we only plot the DOS projected to Fe1 (Os1) atom. DOS was obtained from GGA+U calculations with U = 4 eV for Fe-3$d$ states and U = 2 eV for Os-5$d$ states. The $e_g$-$t_{2g}$ states are indicated for the projected DOS.